\documentclass[%
reprint,
 amsmath,amssymb,
 aps,
]{revtex4-2}

\usepackage{graphicx}% Include figure files
\usepackage{dcolumn}% Align table columns on decimal point
\usepackage{bm}% bold math
\usepackage{xcolor}

\begin{document}

\preprint{APS/123-QED}

\title{Augmented snap-through instability of folded strips}% Force line breaks with \\

\author{Tom Marzin$^{1}$, Barath Venkateswaran$^{1}$, Thomas Baroux$^{1}$, P.-T. Brun$^{1}$ }

\affiliation{$^1$Department of Chemical and Biological Engineering, Princeton University, Princeton, New Jersey 08540, USA}

 \email{pbrun@princeton.edu}

\begin{abstract}

Bistability and snap-through instabilities are central to various mechanisms in nature and engineering, enabling rapid movement and large shape changes with minimal energy input. These phenomena are easily demonstrated by bending a piece of paper into an arch and rotating its edges until snapping occurs. In this Letter, we show that introducing a single crease in such a strip significantly alters its snapping properties. In particular, folded ribbons release much more energy than the regular, unfolded case, leading to faster snapping speeds. %This study explores the physics of this system, where symmetry is broken by plastically deforming the sheet into a minimal origami structure. 
Through numerical simulations and theory, we rationalize our experimental observations.
%, which demonstrate the existence of two distinct snapping regimes. 
We leverage our findings to program the snapping behavior of folded ribbons, demonstrating how our results could find practical applications, e.g., in soft robotics.

%Bistability and snap-through instabilities are central to various mechanisms in nature and engineering. The sudden departure from a stable equilibrium solution to another, possibly distant, solution allows for rapid movement and large shape changes with minimal input. Those problems are classic and can easily be demonstrated by forming an inverted arch with a strip of paper and slowly rotating its edges until it snaps. Here we show that introducing a crease in such a strip dramatically alters its snapping properties, in particular, releasing considerably more energy than in the regular case. In this Letter, we unravel the physics at play in this system where the symmetry has been broken by plastically deforming a sheet to obtain a minimal origami structure. We combine numerical simulations and theory to rationalize our experimental observations, which demonstrate the existence of two distinct snapping regimes. We leverage our finding to program the snapping behavior of such contraptions.
\end{abstract}

%Use showkeys class option if keyword
\maketitle

Slender elastic structures may have two or more equilibrium states, which can be inconvenient when wearing contact lenses or holding an umbrella in a storm. In both cases, the inverted configuration is far less helpful than the natural one. Yet, the transition between these states can be rapid, owing to the snap-through instability that connects them~\cite{liu2021delayed,radisson2023dynamic}. This feature is widely exploited in nature to achieve rapid motion with minimal power resources~\cite{forterre2005venus,smith2011elastic,evans1972jump}. Snapping is also leveraged in engineering with examples including graphene structures \cite{ma2022snap}, microfluidic passive valves \cite{gomez2017passive}, microlens shells \cite{holmes2007snapping} soft jumpers \cite{gorissen2020inflatable}, actuators \cite{overvelde2015amplifying} and robots 
\cite{jin2023ultrafast,zhang2022pneumatically,chi2022snapping}. Folding also offers a mechanism for controlling energy storage and release through complex crease patterns~\cite{brunck2016elastic,li2019architected}, making folded designs ideal for applications requiring compactness, such as in space crafts~\cite{koryo1985method} or medical devices \cite{kuribayashi2006self,rodrigues2017nonlinear}. 
%These structures efficiently manage energy by focusing deformation along folds, improving both functionality and actuation efficiency.\\
In this Letter, we revisit the well-known snap-through instability of a flexible strip \cite{patricio1998elastica,gomez2017critical,radisson2023dynamic} and explore the effect that a localized fold in the strip has on snapping. This minimal alteration of the problem dramatically increases the snapping speed and profoundly modifies the strip stability.  

%The inset highlights the folded region that spans a width $s_0$ much smaller than the length $L$ of the ribbon. Yet, the simple act of folding a ribbon ($\Psi_0 \approx 30^\circ$) 

\begin{figure*}
  \includegraphics[width=1.8\columnwidth]{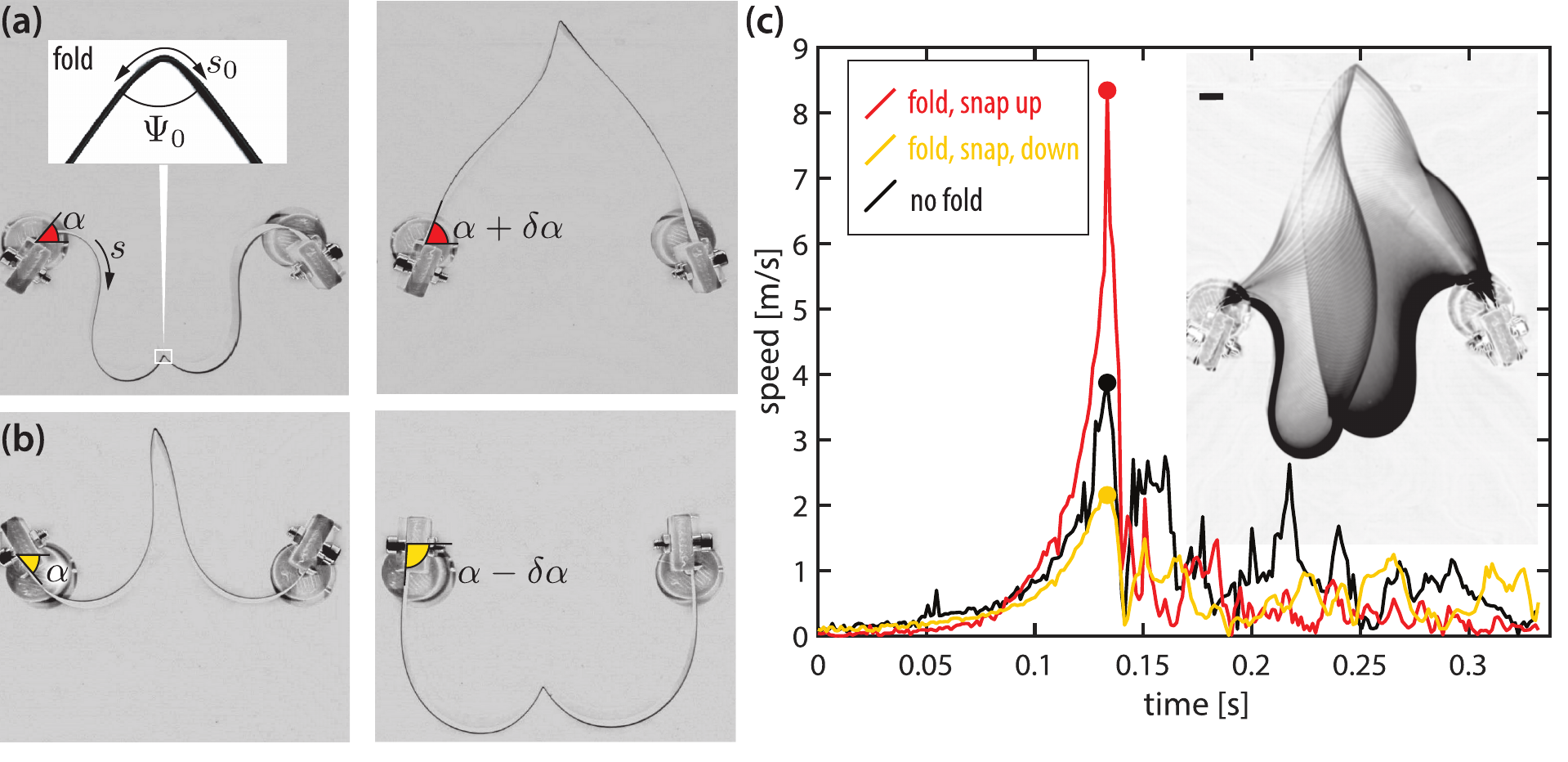}%\columnwidth
  \caption{\textbf{Snapping of a folded strip:} (a) Initial and final configurations when the ribbon snaps up in the direction of the fold.  $(\Psi_0 \simeq 30^{\circ})$ and clamped at both ends (gap $G/L=0.5$) [inset: zoomed-in  view of the localized crease] (b) Reverse snapping down path for the same ribbon. (c) Speed of the fold around the snapping transition for case (a) in  red [inset: chronophotography of the dynamics in (a)] and (b) in yellow. The black line is the reference case without fold, where the mid-point is tracked.}
  \label{fig:1}
\end{figure*}

Folding breaks the symmetry of the problem along the vertical axis. Fig.~\ref{fig:1} shows various strip configurations where the fold always points up. In Fig.~\ref{fig:1}a, the clamp angle $\alpha$ slowly increases, and the ribbon moves up, while in Fig.~\ref{fig:1}b, $\alpha$ is decreased, and the ribbon moves down. The transition between those pairs of configurations is rapid, as expected from a snap-through instability. 
In Fig.~\ref{fig:1}c, we report experimental results for a particular case where the fold reaches $8\,$m/s (see the red curve), while a similar strip \textit{without} fold merely reaches $4\,$ m/s (see black curve). This augmented speed is caused by the fold whose width $s_0$ is much smaller than the length $L$ of the ribbon (see inset of Fig.~\ref{fig:1}a). In this Letter, we characterize and study the snapping dynamics illustrated in the chronophotography in Fig.~\ref{fig:1}a and rationalize the augmentation of the snapping speed with folding.

%where the symmetry of the kinematic of snapping transition along the horizontal direction appears broken.\\

 %Next, we focus on the dynamics of the snapping process. Finally, we discuss potential applications of this mechanism as a fast and selective actuator.

\begin{figure}
  \includegraphics[width=\columnwidth]{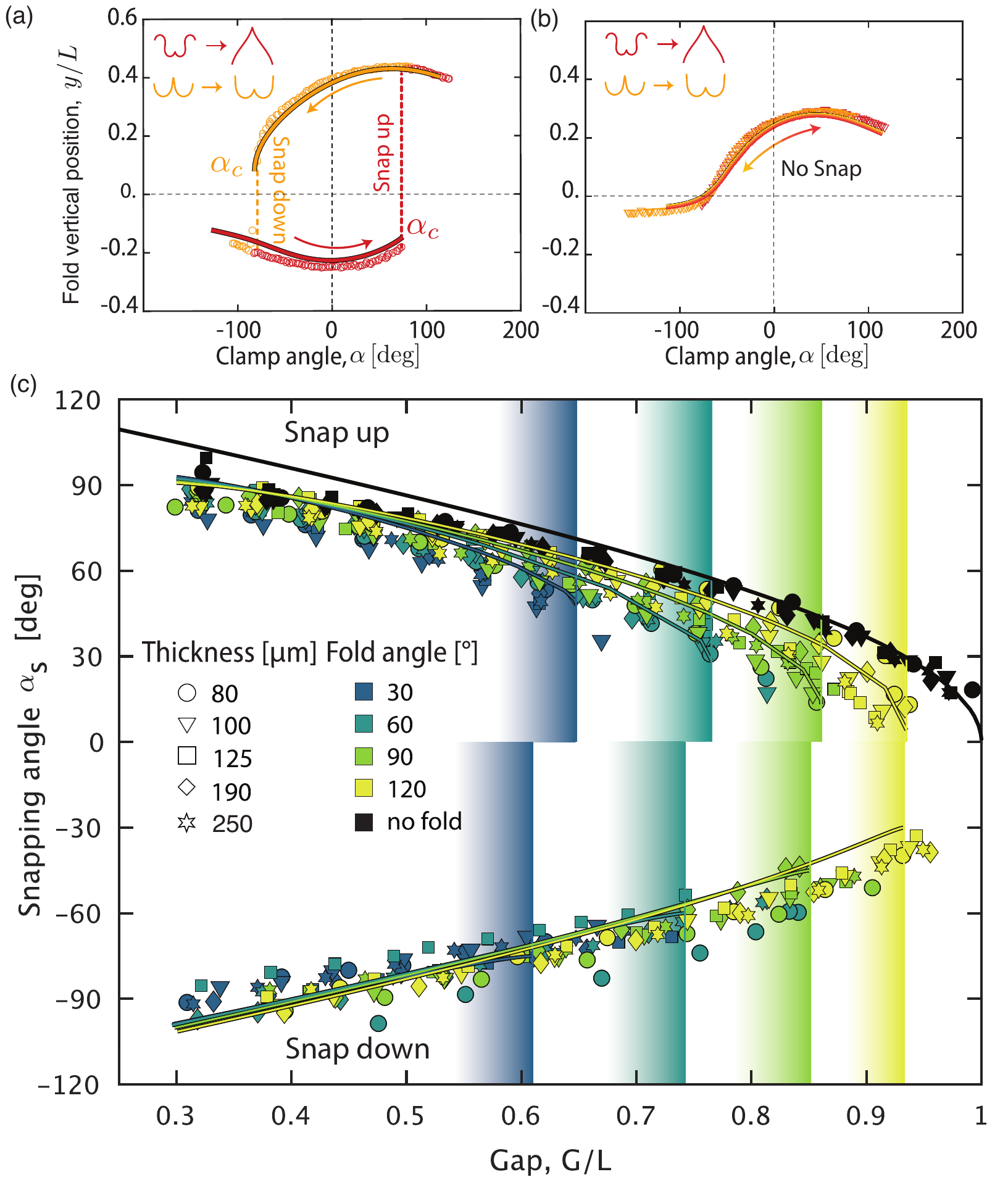}%\columnwidth
  \caption{\textbf{Stability analysis:} Stable quasi-static fold ($\Psi_0 \approx 60^\circ$) vertical position against the clamp angles for both paths (marker color) for two different gaps respectively (a) $G/L=0.5$ and (b) $G/L=0.8$. (c) Critical clamp angle, when the snap occurs as a function of the resealed gap. Each half part of the graph corresponds to one path: snap up ($\alpha_c>0$) and snap down ($\alpha_c<0$) respectively. The vertical lines are the theoretical limit of the snapping domain for each fold angle. Error bars (omitted) from the repetition of the same experiments rarely exceed the symbol size. For all graphs, the solid lines are Kirchhoff equations' solutions at respective snapping limits}
\label{fig:2}
\end{figure}

%We first examine how folding a ribbon affects the snapping instability. 
 We begin by examining how the fold and its properties affect the stability of the strip. In Fig.~\ref{fig:2} and Movie S1, we report the results obtained with thin polyethylene terephthalate strips of length $L = 20 \,\text{cm}$ and $\Psi_0 \approx60^{\circ}$. We clamp the ends of the sample strips and symmetrically rotate them using stepper motors. Note that the effect of the clamps' rotation speed on the instability is minimal in the parameter space we explored (see SI).  In Fig.~\ref{fig:2}a, we show how, in a typical snapping experiment, the vertical position of a fold varies with changing clamp angle $\alpha$.  For the dimensionless gap length, $G/L = 0.5$, we observe continuous variations of the shape until a critical value $\alpha_c$, after which a large discontinuous jump occurs. This rapid change signals the snapping transition. Snapping up (in the direction where the fold points) is observed for $\alpha_c \approx 73.6^{\circ}$. The opposite snapping path shows a smoother transition with a smaller jump with $\alpha_c \approx -80^{\circ}$. The resulting hysteresis cycle is characteristic of the bistability of strips~\cite{gomez2017critical}, albeit its asymmetry is unique to the folded case. Even more surprising is that snapping and, thus, hysteresis disappear beyond specific gap lengths. Fig.~\ref{fig:2}b shows that the fold's vertical position smoothly varies with the clamp angle in both directions for  $G/L=0.8$ and $\Psi_0=30^\circ$. This result is unique to folded strips, as their unfolded counterparts always snap.
In Fig.~\ref{fig:2}c, the sheet thickness, $t$, is varied from 80 to 250 $\mu$m to avoid plasticity, while the crease angle, $\Psi_0$,  is set to $30^\circ$, $60^\circ$, $90^\circ$, $120^\circ$ ($\pm 10^\circ$, see fold manufacturing details in SI). The no-fold case was added as a point of reference. We plot the snap angles, $\alpha_c$ against the dimensionless gap $G/L$ for both snapping directions. The data across thicknesses collapses onto master curves that solely depend on $\Psi_0$, indicating that the magnitude of the snap angle is a decreasing value of the gap, for a given crease angle. These curves also end abruptly as snapping disappears beyond a threshold value of $G/L$, giving way to smooth transitions between stable configurations (see Fig.~\ref{fig:2}b). The end of the snapping branches appears to be a function of the fold angle alone. Lower values of $\Psi_0$, i.e., tighter folds, have their master curves terminating sooner than wider folds. The limiting case of $\Psi_0 = 180^\circ$ (unfolded strip) does not exhibit termination.
%\pt{Finally, we note that snapping sooner for negative values of $\alpha_c$ than for the positive ones.} \pt{remove??}
 \newpage

To rationalize these results, we model our system using the Kirchhoff equations for elastic rods~\cite{audoly2000elasticity}:
\begin{align}
\label{eq:kir}
    \mathbf{f}'=\lambda\ddot{\mathbf{r}} \nonumber\\
    \mathbf{m}' + \mathbf{r}' \times \mathbf{f} =\mathbf{0} \nonumber\\
    \mathbf{m} = B \left(\mathbf{\theta'}-\mathbf{\theta}_0'\right)\mathbf{e}_z 
\end{align}
where $\mathbf{r(s,t)}$ denotes the position of the beam as a function of time $t$ and arclength $s$ (Fig.\ref{fig:1}a), dots and primes denote time derivatives and derivatives along the arclength. The orientation of the ribbon is $\theta(s)$, such that the tangent to the ribbon is $\mathbf{r}'=[\cos(\theta), \sin(\theta)]$. 
The internal force and moment are denoted $\mathbf{f}$ and $\mathbf{m}$. The ribbon has a lineic mass $\lambda$, bending stiffness $B$, and natural curvature $\theta_0'(s)$. We model the crease with a step-like function~\cite{lechenault2014mechanical, jules2019local}:
\begin{equation}
\label{eq:const}
    \theta_0(s)  = \frac{\Psi_0-\pi}{2}\tanh \left( \frac{s-L/2}{s_0}\right)    
\end{equation}
This formulation accounts for the fold angle $\Psi_0$, the fold width $s_0$, and its position at the center of the beam, $L/2$. 

The steady-state case ($\ddot{\mathbf{r}} = 0$) is solved using a continuation method, implemented in \texttt{AUTO-07p}~\cite{doedel1998auto} (see SI for details). The stable branches  show excellent agreement with experiments, accurately capturing their variation with $\alpha$ with or without snapping (see Fig.~\ref{fig:2}a-b). The solid lines in Fig.~\ref{fig:2}c represent the snapping angles $\alpha_c$ for different fold angles, as predicted by our model. We find a fair agreement between data and theory, even in predicting the branch termination. The overall quality of the results suggests that our model accurately captures the physics of the system. We now move the study to the dynamics at play, aiming to rationalize the increase in snapping speed enabled by the mere presence of a fold. 

\begin{figure}
  \includegraphics[width=\columnwidth]{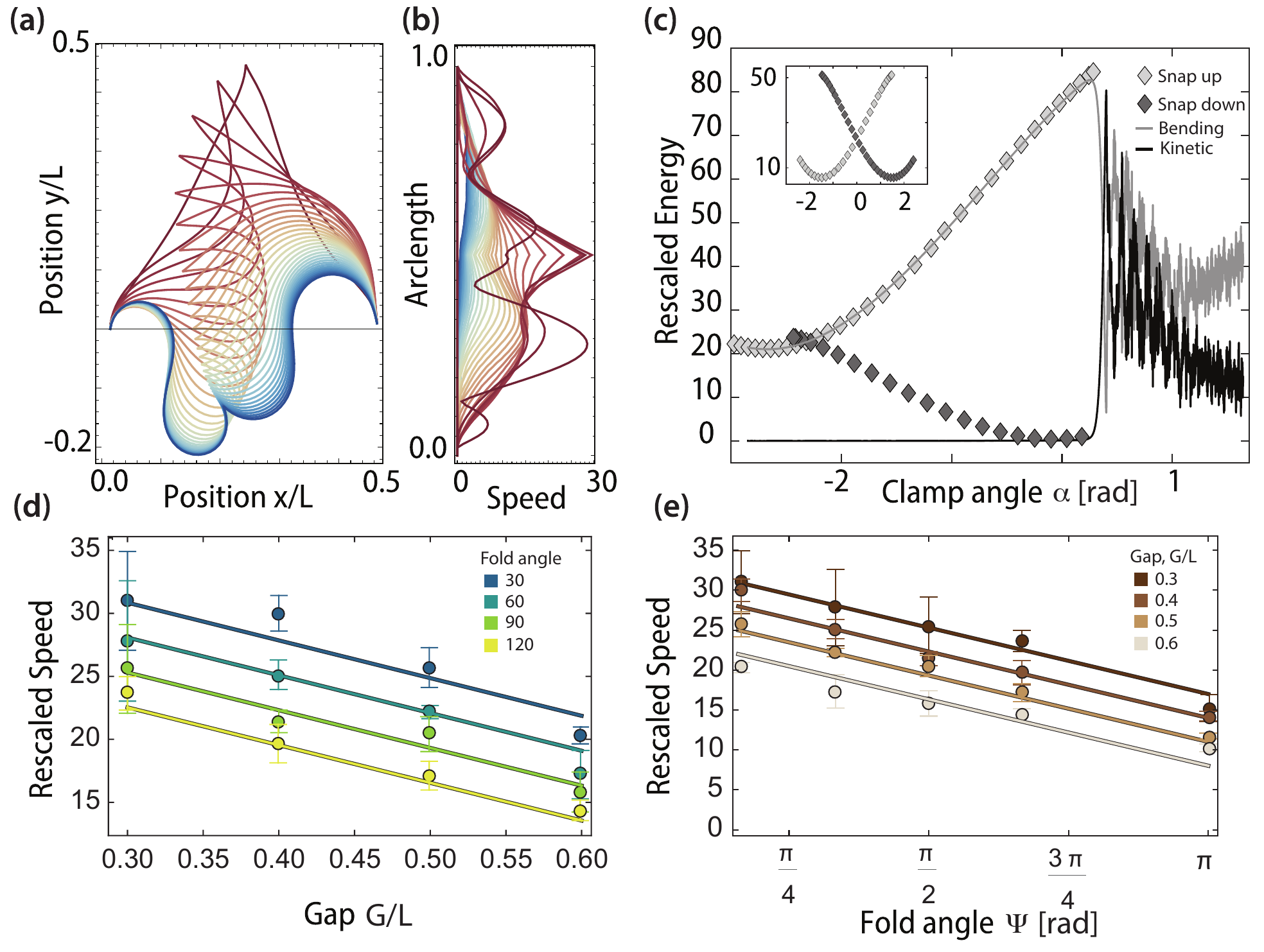}%\columnwidt}
  \caption{\textbf{Dynamics of snapping:} (a) Dynamics of snapping as captured by DER and (b) corresponding speed distribution along the folded ribbon   ($\Psi_0 = 30^{\circ}$ and $G/L = 0.5$). (c) Evolution of the bending and kinetic energies in DER simulations (solid lines). Symbols denote the bending energy of both equilibrium configurations as computed by \texttt{AUTO-07p} (inset: no fold case). (d) (resp. e) Snapping speed as a function of the gap (resp. fold angle). Lines correspond to the plane in Eq.~\eqref{eq:plane}.
}
\label{fig:3}
\end{figure}

In Fig.~\ref{fig:3}a, we show the snapping dynamics as predicted when solving Eqns.~\eqref{eq:kir} and~\eqref{eq:const} using the discrete elastic rod (DER) algorithm~\cite{bergou2008discrete}. The results of the simulations compare favorably with experiments without any adjustable parameters (see the inset of Fig.~\ref{fig:1}c as a comparison). In particular, we recover the asymmetry of the snapping motion, showing that the snapping of a folded beam can be treated as two connected beams, one for $0\leq s< L/2$ and another one for $L/2<s\leq1$.  In Fig.~\ref{fig:3}b we show that the second beam snaps first as positive values of the speed are recorded for large values of $s$ only. This motion then entrains the second snapping where larger values of the speed are recorded. For a while, the fastest point is $s\simeq L/4$, but eventually, the fold ($s=L/2$) achieves the greatest speed. We note that the fold does not bend and that the curvature of this region remains unchanged throughout the motion. In fact, our ribbons are longer than the critical origami length $L^*\simeq 1.6-5$~cm, obtained by balancing the bending facets and fold stiffnesses \cite{lechenault2014mechanical}. This length is the cutoff between the rigid facet regime~\cite{marzin2022shape,nain2024tunable} and the rigid fold regime explored in this study. In particular, the crease remains intact after several consecutive actuations (see SI).

In Fig.~\ref{fig:3}c, we report the beam's bending and kinetic energy as they evolve when the clamp angle is modified. Unsurprisingly, we note a smooth evolution of these quantities, where the kinetic energy remains virtually nil, and the bending energy follows the prediction of static calculations up to the point where snapping occurs. At this point, the bending energy is almost entirely traded off for kinetic energy. The curves cross each other, and the kinetic energy reaches the value of the bending energy right before snapping. 
Introducing a fold skews the energy landscape (see inset of Fig.~\ref{fig:3}c for a symmetric energy landscape present in the unfolded case), thereby increasing the energy gap between stable configurations.  Once released, this increased bending energy allows for greater snapping speeds. Leveraging this understanding, we anticipate that the snapping speed $v_s$ scales with the speed gauge resulting from the balance of bending and kinetic energy $v^*=\sqrt{B/\lambda L^2}$. We find that $\bar{v}_s=v_s/v^*$ is a function of geometry alone. 
In Fig.~\ref{fig:3}d, we show that $\bar{v}_s$ depends linearly on the dimensionless gap $G/L$ for a fixed value of the fold angle. In Fig.~\ref{fig:3}e, we show that $\bar{v}_s$ is an affine function of the fold angle $\Psi_0$ for a fixed value of the gap $G/L$. Overall, $\bar{v}_s$ is described by the plane:
\begin{equation}
\label{eq:plane}
   \bar{v}_s= \bar{v}_m - a \, G/L- b \, \Psi_0
\end{equation}
with $\bar{v}_m=42.59\pm1.33$ being the order of magnitude of the rescaled snapping speed, and $a=30.38 \pm 2.57$ and $b=5.00 \pm 0.26$ being the coefficients obtained from fitting a plane to experimental data (see the projections in Fig.~\ref{fig:3}c-d). 

%Note that the effect of the clamps' rotation speed on the instability is minimal in the parameter space we explored (see SI). 

\begin{figure*}
  \includegraphics[width=\textwidth]{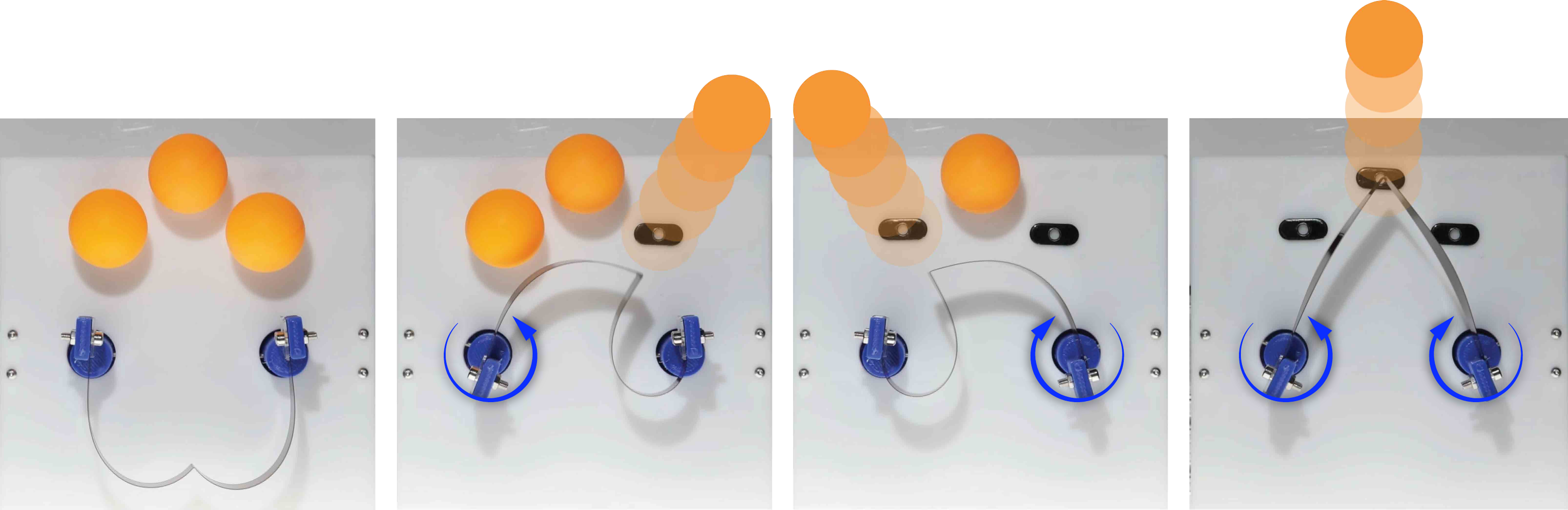}%\columnwidth
  \caption{\textbf{Folded strip as a fast and programmable actuator:} Sequential actuation of the ribbon enables us to selectively impact objects.}
  \label{fig:application}
\end{figure*}

We now discuss the merits and limitations of Eqn.~\eqref{eq:plane}. The shapes of the ribbons before snapping are equilibrium solutions matching Euler's fifth and sixth species~\cite{truesdell1960rational}. In the folded case, two elastica are needed to build the total shape (see SI). The typical dimensionless bending energy of these units is around $100 B/L$, yielding typical speeds of $10\,v^*$ when equating bending and kinetic energies. This value corresponds to the mean value of the beam speed when snapping, which agrees well with the data in Fig.~\ref{fig:3}a, and the order of magnitude of  $\bar{v}_m$ in Eqn.~\eqref{eq:plane}. Nevertheless, we note that $\bar{v}_m$ is not attainable in experiments since it requires $G/L=0$, which would lead to self-contact during snapping. In practice, we are limited to around $75 \%$ of $\bar{v}_m$ before self-contact occurs.
Further examination of the Elastica solutions explains the linear dependence of the speed in $G/L$. Increasing the gap yields a quadratic decrease in bending energy, and hence, the linear decline in speed seen in Eqn.~\eqref{eq:plane}. Likewise, noticing that folded ribbons effectively comprise two Elastica matched at the location of the fold allows us to infer that the bending energy varies quadratically with $\Psi_0$ in the range explored. As such, the speed varies linearly, as seen in Eqn.~\eqref{eq:plane} (see SI for more details on the energy landscape of the Elastica).

In this Letter, we demonstrated that introducing a localized fold into a strip significantly alters the snapping transitions. This plastically induced localized curvature allows the strip to store more energy when bent than its flat counterpart. When this energy is released, greater snapping speeds are recorded. While the order of magnitude of these speeds and their dependence on key parameters can be rationalized using energy-based arguments, we note that the exact value of the maximal speed depends on the complex dynamics that unfold.

%  The size of the snapping domain is directly influenced by the fold angle. By solving the Kirchhoff equations, we can predict the experimental results and anticipate the asymmetric deployment of the strip during the transition, as well as the typical speed of the ribbon. Notably, we show that a sharper angle can induce faster motion for a given gap between the clamps.\\
% However, this system has certain limitations. Self-intersection is a concern when the clamps are too close, as parts of the ribbon may touch, causing friction, energy loss, and reduced snapping speed (see SI). 

This dynamic behavior can be applied in practical settings. Inspired by the works of Refs.~\cite{liu2021delayed,radisson2023dynamic} we impose asymmetric boundary conditions in our problem. In Fig.~\ref{fig:application} and Movie S2, we show how snapping can be selectively controlled in our folded ribbons. First, the right clamp is kept at a fixed angle while we rotate the left clamp. This choice only induces snapping in the left part of the ribbon, propelling the right-most ball in Fig.~\ref{fig:application}; but leaving the other ones untouched. Reversing the dynamics allows us to remove the left-most obstacle, leaving the obstacle in the center untouched. Finally, actuating both clamps simultaneously helps push this last obstacle away. 
This experiment demonstrates the gain in functionality that the folds enabled, showcasing the potential of folded strips as fast, selective actuators, e.g., for soft robotics systems. Beyond soft robotics, folded ribbons could also find applications wherever precise and swift actuation in confined spaces is needed~\cite{kuribayashi2006self,kim2013soft}. \\
Our approach also connects to origami, where folding is central to this practice, and is now successfully applied in various engineering domains, from deployable solar panels~\cite{koryo1985method} to medical devices~\cite{kuribayashi2006self}. While multi-stable, origami-inspired structures have been extensively explored under diverse loading scenarios~\cite{feng2018twist,gori2022deployment,zhang2023lead,marzin2022shape, nain2024tunable}, dynamical effects are seldom considered. Our study suggests that carefully located folds could allow origami to be rapid, energy-efficient actuators.

%\cite{lechenault2015generic,flores2022effect}

%Moreover, the snap-through mechanism could be leveraged for energy harvesting systems, where the ribbon's ability to store and quickly release energy offers promising potential in self-powered sensors and wearable technologies \cite{tang2018leveraging,wang2015flexible}.

%\textbf{Add something about origami.}

%OLD Tom (before PT's suggestions)

%The transition between two stable states can also result in significant morphological changes. This principle is fundamental to folded structures which often involve multiple stable states through the crease patterns \cite{brunck2016elastic,li2019architected}. Such structures offer advantages in compaction and deployment of devices, such as solar panels \cite{koryo1985method} or tubular structures for medical applications \cite{kuribayashi2006self,rodrigues2017nonlinear}, while involving minimal energy needed to actuate by localizing the deformation along the fold. 

\begin{acknowledgments}
We wish to acknowledge Lauren Dreier, Yuchen Xi, Julien Le Dreff, and Romain David's helpful advice and discussion in improving the experimental set-up.
PTB and TM acknowledge the financial support from the Princeton Innovation funds. 

\end{acknowledgments}
\appendix

\end{document}